# DNA-Origami-Assembled Rhodium Nanoantennas for Deep-UV Label-Free Single-Protein Detection

*Nicco Corduri[1], Malavika Kayyil Veedu[2], Yifan Yu[3], Yanqiu Zou[4], Jie Liu[3], Denis Garoli[4,5,6*], Guillermo P. Acuna[1,7*], Jérôme Wenger[2*], Karol Kołątaj[1*]*

[1] Department of Physics, University of Fribourg, Chemin du Musée 3, Fribourg CH-1700, Switzerland.

[2] Aix Marseille Univ, CNRS, Centrale Med, Institut Fresnel, AMUTech, 13013 Marseille, France.

[3] Department of Chemistry, Duke University, Durham, NC 27708, USA.

[4] College of Optical and Electronic Technology, China Jiliang University, Hangzhou 310018, China

[5] Optoelectronics, Istituto Italiano di Tecnologia, 16163 Genova, Italy.

[6] Dipartimento di Scienze e Metodi dell'Ingegneria, Università di Modena e Reggio Emilia, Via Amendola 2 Padiglione Tamburini, 42122 Reggio Emilia, Italy.

[7] Swiss National Center for Competence in Research (NCCR) Bio-inspired Materials, University of Fribourg, Chemin des Verdiers 4, CH-1700 Fribourg, Switzerland.

**Corresponding authors:** denis.garoli@unimore.it, guillermo.acuna@unifr.ch, jerome.wenger@fresnel.fr, karol.kolataj@unifr.ch

**Abstract**

Plasmonic metal nanoparticles have deeply impacted the spectroscopy field by enabling nanoscale concentration of light and powerful signal enhancement. However, their operation remains largely confined to the visible and near-infrared spectral ranges due to the poor stability and limited surface functionalization strategies of UV-active materials. Here, we overcome these barriers and establish a complete route for the fabrication of programmable UV-plasmonic nanoantennas based on rhodium nanocube dimers assembled on DNA origami scaffolds. We report on an effective surface ligand exchange protocol for functionalizing rhodium nanocubes with DNA strands without compromising the colloidal integrity of the nanoparticles. This turns the rhodium nanocubes into buildings blocks, which we assemble into dimers with a yield of 69%, achieving an average interparticle distance of 10 nm. Thanks to the DNA origami, a single streptavidin protein is accurately positioned into the plasmonic nanogap, contrasting to other works relying only on random diffusion. In good agreement with numerical simulations, UV autofluorescence measurements on single streptavidins indicate brightness enhancement factors up to 22×, shorter autofluorescence lifetimes, and over 10× increase of the total photon budget. By pioneering a robust, versatile and selective strategy for constructing UV-resonant plasmonic nanoantennas, this work broadens the applications of plasmonics into the deep UV range and opens new opportunities for label-free spectroscopy of single proteins.

## Introduction

The ability of plasmonic metal nanoparticles (NPs) to concentrate the electromagnetic field at the nanoscale has been widely exploited for enhanced spectroscopy techniques such as surface-enhanced Raman scattering (SERS), fluorescence, and circular dichroism.[1–3] However, unlocking the optical potential of plasmonic NPs into practical plasmonic-enhanced spectroscopy tools requires a careful consideration for the NP assembly and their surface functionalization.[4,5] A particularly effective strategy is to graft the surface of NPs with a monolayer of thiol-terminated single-stranded DNA (ssDNA) forming strong sulfur bonds to the metal.[6–8] Such DNA shells not only stabilize the NPs in high-salt media but also encode sequence-specific recognition, transforming the NPs into uniquely addressable, programmable building blocks.[8] DNA origami is a highly adaptable bottom-up technique that makes full use of DNA's programmable nature, enabling the precise placement of nanoparticles and other components (such as fluorophores, quantum dots, proteins, enzymes, or nucleic acids) with nanometer-level accuracy and exact stoichiometric control down to individual molecules[9,10] This deterministic placement fully leverages the optical properties of plasmonic NPs and enables reproducible enhancement of fluorescence and SERS down to the single molecule level.[11,12]

To date, DNA functionalization has been realized predominantly for Au and Ag NPs with some excursions into high refractive index materials[13,14]. Thus, DNA and DNA origami–based plasmonic systems operate almost exclusively in the visible/near-IR.[9,11] However, this spectral range misses the many opportunities open in the ultraviolet (UV) where molecular absorption is the highest, notably the π–π* bands of aromatic amino acid residues such as tryptophan.[15–22] Extending the operating range of origami-based plasmonic nanoantennas towards the deep-UV unlocks the key benefit to probe directly the protein intrinsic UV autofluorescence and avoids the need for additional fluorescence labeling.[23–31] Moreover, the deep UV range enables the exploitation of resonance enhancement in SERS, where spectral overlap between the plasmonic mode and molecular absorption bands leads to significantly amplified signals.[21,32–37] DNA origami-based nanoantennas provide the supplementary crucial advantage of accurately localizing the protein target inside the antenna nanogap hotspot with nanometer and stoichiometric control.[10,38–40] So far, the earlier works on UV plasmonic antennas relied on nonspecific deposition of dense molecular layers,[32–35,41–46] or random translational diffusion into the hotspot area,[31,47,48] but DNA origami antennas and their high potential to deterministically locate a single target emitter into the hotspot remain unexplored in the deep UV range.

Beyond the visible, a subset of materials, most notably Al, Ga, Mg, and Rh, supports localized surface plasmon resonances in the UV range.[17,20,49–52] However, despite a decade of synthetic advances that yield diverse shapes and sizes of UV-active NPs, their practical application remains strongly limited by their surface chemistry and stability.[53–56] Al, Mg, and Ga are highly reactive, with large oxidation enthalpies.[49] While stable in organic solvents, they therefore tend to oxidize in aqueous environments.[53,57] Their stability could be increased by depositing a dielectric coating,[53,58] yet this coating substantially dampens the near field intensity enhancement and reduces the spectroscopic performance.[59,60] To overcome these limitations, rhodium stands out as a particularly promising material for UV plasmonics combining strong UV resonances with high chemical stability.[61–67] However, most of the UV-active Rh NPs such as nanocubes (NC) or tripods are synthesized using organic polyol routes in the presence of polyvinylpyrrolidone (PVP) molecules. This ligand protects the growing crystals and steers their morphology to ensure high reproducibility of their size and shape.[61–63] At the same time, PVP binds so efficiently and densely to the metal surface that it blocks its surface to the interaction with other molecules, thus hindering the functionalization. Consequently, the colloidal functionalization of Rh NPs with DNA has not been achieved so far. Establishing a robust method of removing PVP from Rh surface and their subsequent DNA functionalization would unlock the fabrication route of programmable UV nanoantennas with deterministic analyte placement, expanding plasmon-enhanced fluorescence and resonant SERS into the deep-UV regime where light-molecule interaction is the strongest.

In this work, we pioneer a route for integrating UV-active plasmonics together with DNA origami nanotechnology. Our first major result is a reliable strategy for functionalizing Rh NCs with DNA strands, which involves a dedicated approach to remove the PVP ligands from the Rh surface. Second, using the DNA origami scaffold, we assemble well-defined UV-resonant Rh dimer antennas in a face-to-face configuration, achieving an average interparticle distance of 10 nm and a yield of 69%. A single streptavidin protein is deterministically positioned into the plasmonic nanogap, contrasting to earlier works relying only on random diffusion.[30,47,48] The UV autofluorescence intensity recorded from the streptavidin protein inside the hotspot indicate an enhancement up to 22× as compared to the reference on quartz, together with a 6.6× reduction of the autofluorescence lifetime and a 10× increase of the total photon budget. The experimental results are in very good agreement with numerical simulations, highlighting the high nanofabrication and positioning accuracy enabled by our UV-active nanoantennas. By establishing a robust and versatile strategy for the realization of UV-resonant Rh nanoantennas, our work broadens the applicability of plasmonics down into the deep UV range and paves the way toward enhanced label-free spectroscopy with single molecule resolution.[16,23,68,69]

**Results and discussion**

Rhodium NCs with a side length of 23.6 ± 1.9 nm are selected as a model UV-plasmonic system for their strong localized surface plasmon resonances (LSPRs) in the UV range and their excellent chemical stability in aqueous environments.[63] Transmission electron microscopy (TEM) images shown in the Supplementary Fig. S1a,b confirm the successful synthesis using the polyol method (see Methods section for details about the Rh NCs synthesis protocol).

To construct programmable UV-plasmonic Rh nanoantennas on DNA origami scaffolds, we need to establish an efficient ligand-exchange process that preserves colloidal stability and enables functionalization of the Rh surface with DNA. Our dedicated surface modification protocol involves two steps: first, we remove the PVP ligand bound to the metal surface, and then we introduce thiolated single-stranded DNA (ssDNA), as shown schematically in Fig. 1a. The PVP binding mechanism is based on the formation of coordination bonds between the Rh atoms and the carbonyl oxygen of the PVP molecules (Fig. 1b). To expose the metal surface for interaction with DNA, we use sodium borohydride ($NaBH_4$) to reduce the coordination bonds and compete with the PVP molecules, removing them from the NCs. However, $NaBH_4$ is a strong reducing agent that can destabilize colloids and cause their aggregation. Therefore, the process was carefully optimized to be carried out at low temperature, assisted by sonication to minimize aggregation, and then stabilized by adding the competing surfactant (Pluronic F-127) to replace the detached PVP molecules (see the Materials and Methods section for details). After PVP removal, the NCs are functionalized with thiolated poly-thymine (poly(T)18) DNA strands routinely used for the functionalization of metallic NPs. Functionalization is performed using a "Freeze–thaw" method, in which the freezing step promotes efficient binding between the metal and thiolated DNA, as shown in Fig. 1c.[70]

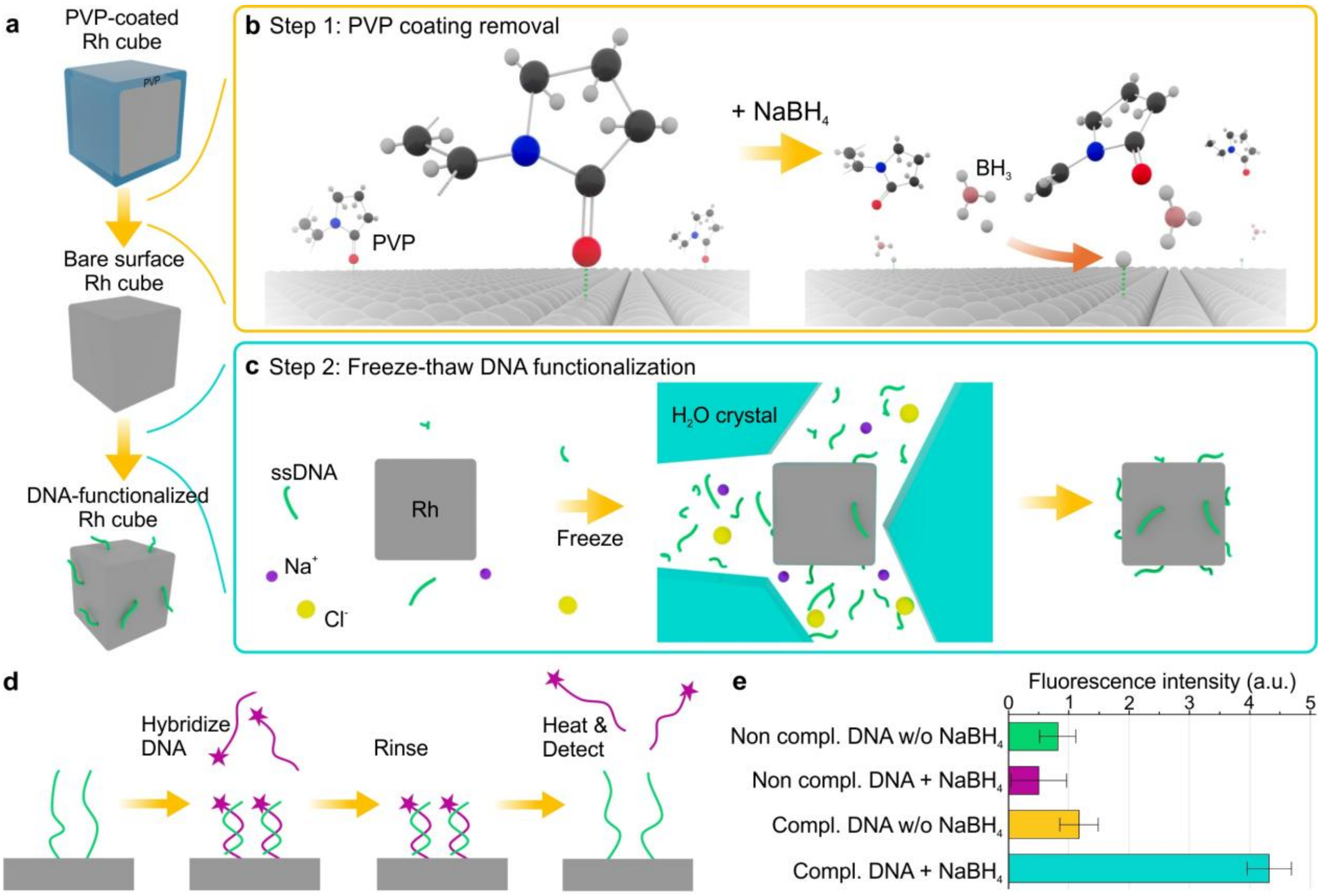

**Figure 1.** Surface functionalization of rhodium NCs with single stranded DNA. (a) Full workflow of the DNA functionalization of Rh NCs. (b) Mechanism of PVP removal from Rh NCs by $NaBH_4$ through cleaving Rh-carbonyl bond. (c) Schematic representation of the NC surface functionalization with DNA using the "Freeze–thaw" method. (d) Workflow of the fluorescence assay used to quantify DNA attachment on the Rh NCs. (e) Fluorescence intensity measurements for Rh NCs with and without PVP coating incubated with either a complementary or a non-complementary DNA strand labeled with the fluorescent dye ATTO 647N.

To assess the performance of our Rh surface functionalization protocol, we use a fluorescence assay whose principle is summarized in Fig. 1d. The assay relies on the specific hybridization of the fluorophore-labeled DNA strands (Poly(A)18 - ATTO 647N) complementary to the poly(T)18 DNA strands on the NC surface. After hybridization and washing, the fluorescence from the surface-hybridized strands is measured. A strong fluorescence signal is observed only for the complementary DNA strands using the NCs treated with $NaBH_4$ (Fig. 1e). On the contrary, untreated Rh NCs samples retaining the surface PVP ligands show only weak nonspecific emission comparable to that observed when a non-complementary DNA sequence is used. These results confirm the successful surface functionalization of Rh NCs with specific thiolated DNA strands, highlighting that PVP removal represents a critical step to enable DNA functionalization. From the fluorescence intensity in Fig. 1e, we estimate the surface density of DNA to be approximately 0.05 molecules $nm^{-2}$, consistent with the densities reported for Ag NCs.[71] Gel electrophoresis confirms our results (Supplementary Fig. S1d), where only NCs functionalized after PVP removal migrated through the gel, indicating successful DNA conjugation and increased surface charge. Additionally, as can be seen from the gel analysis in the Supplementary Figure S1d and e, DNA functionalization combined with gel electrophoresis allows for the separation between the different fractions formed during the NCs synthesis, which is another crucial element for the optical applications.

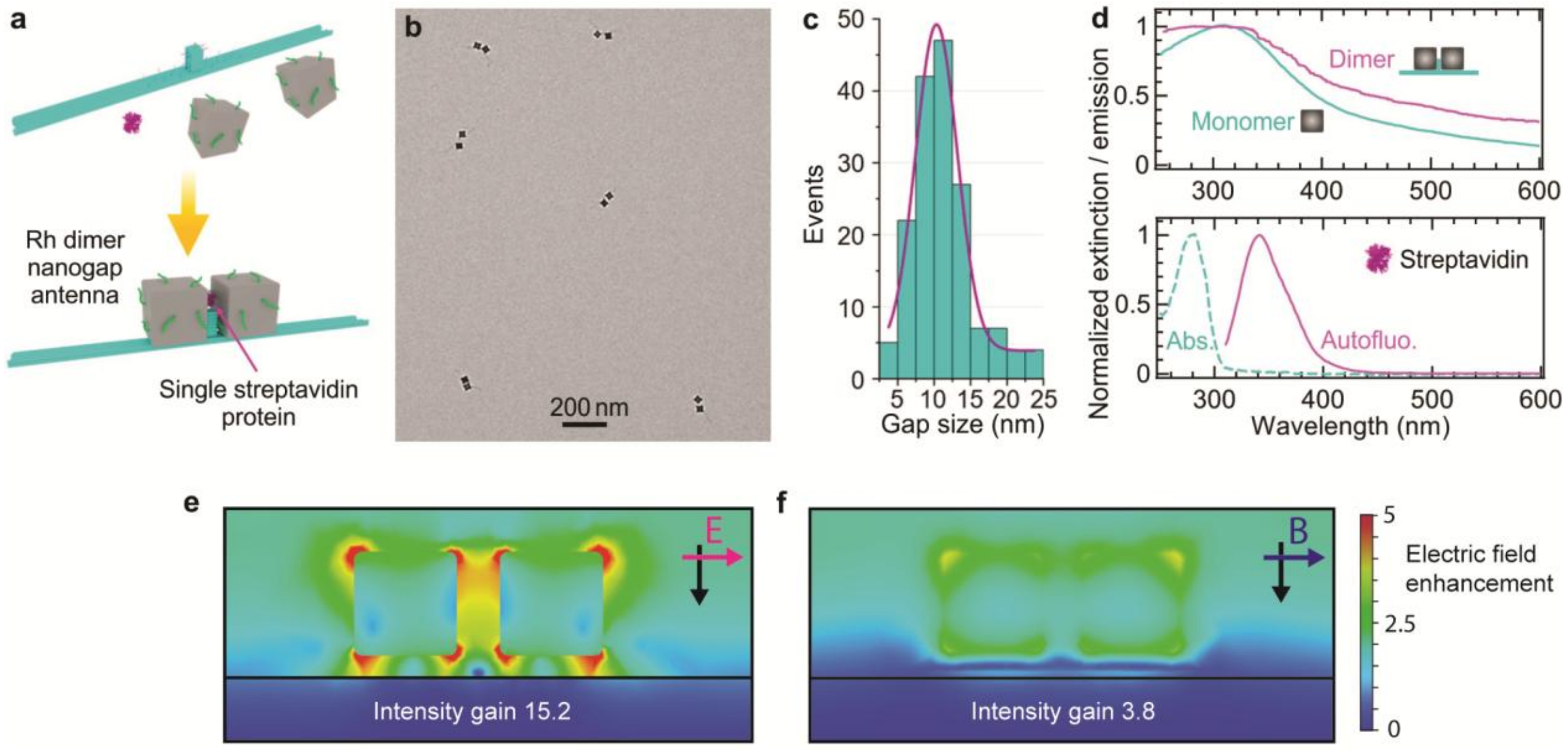

**Figure 2.** Deep-UV self-assembled plasmonic nanoantennas. (a) Realization of a UV nanoantenna composed of a Rh NC dimer with a single streptavidin precisely positioned in the hotspot via a DNA origami scaffold. (b) TEM image of 24 nm Rh NC dimer antennas. (c) Gap size distribution obtained from TEM images out of 164 individual dimer antennas. (d) Normalized extinction spectra of the Rh dimer antenna and single Rh NCs, together with the normalized absorbance and autofluorescence emission of a 12 µM streptavidin solution. (e, f) Distribution of the electric field enhancement around a rhodium dimer nanoantenna with 10 nm gap under linearly polarized light at 266 nm with the electric field (e) parallel and (f) perpendicular to the dimer main axis. The intensity gain indicated at the bottom corresponds to the streptavidin position (black dot marker).

Having achieved a reliable surface functionalization of the Rh NCs with DNA, we now use DNA origami scaffolds to assemble the NCs into well-defined plasmonic nanogap antennas designed to enhance the UV autofluorescence of native proteins (Fig. 2a). The rectangular base of our DNA origami template features a length of 180 nm, a width of 15 nm and a height of 5 nm. It includes a “mast” of 8.5 nm width and 20nm height placed in the center (Supplementary Fig. S2).[14,72] A single biotin extension is added at the top of the mast to precisely bind a single streptavidin protein and locate it inside the plasmonic hotspot. We check the binding of streptavidin to the DNA origami by fluorescence microscopy as shown in Supplementary Fig. S3. To attach the Rh NCs to the DNA scaffold, the origami features 32 poly(A)8 handles complementary to the poly(T)18 DNA on the NC surface. Half of the hybridization handles protrude from the base, with the other half protruding from the mast. This design ensures specific and reproducible Rh NC dimer formation while leaving the hotspot area accessible to the target molecule.[73] Transmission electron microscopy confirm the formation of well-aligned dimers with an average interparticle distance of 10.7 ± 3.2 nm and a yield of 69% (Fig. 2b, c).

The extinction spectrum of the Rh NCs dimer shows a plasmonic resonance at 320 nm, slightly red-shifted as compared to the plasmon band of the Rh NC monomer (Fig. 2d). Because of the random orientation of the surfaced-immobilized dimer nanoantennas and the lack of polarization control, we cannot resolve the longitudinal and the transversal resonances of the dimer.[74] Importantly, the plasmon resonance of the dimers still overlaps with the absorption and emission band of streptavidin in the UV (Fig. 2d), indicating the strong UV activity of Rh NCs make them relevant candidates for UV-enhanced spectroscopy.

To evaluate the antenna near-field distribution under 266 nm illumination, consistent with our experimental excitation wavelength, we perform numerical simulations on a Rh dimer with 24 nm side length and 10 nm gap corresponding to the experimental conditions (Fig. 2e, f). The Rh NCs antenna is placed above an Al mirror to further increase the excitation and emission gains,[75,76] with a 5.5 nm spacing corresponding to the thickness of the DNA origami. The near-field amplitude maps in Fig. 2e, f show that the dimer behaves like an efficient optical nanogap antenna, concentrating the electric field inside the gap and demonstrating a polarization-sensitive response. When the incoming field polarization aligns with the dimer main axis (Fig. 2e), the plasmonic coupling leads to a strong electromagnetic confinement with a 15.2× amplification of the local intensity. When the illumination polarization is rotated by 90° (Fig. 2f), the coupling decreases and the intensity gain is reduced by approximately 4 times. We also tested numerically whether small angular misalignments of the NCs in the dimer could affect the antenna’s performance. To our surprise, the simulations shown in the Supplementary Fig. S4 indicate a moderate impact of NCs angling on the near-field strength and emission intensity, indicating that our UV nanoantennas are quite robust to minor geometric deviations without loss of performance.

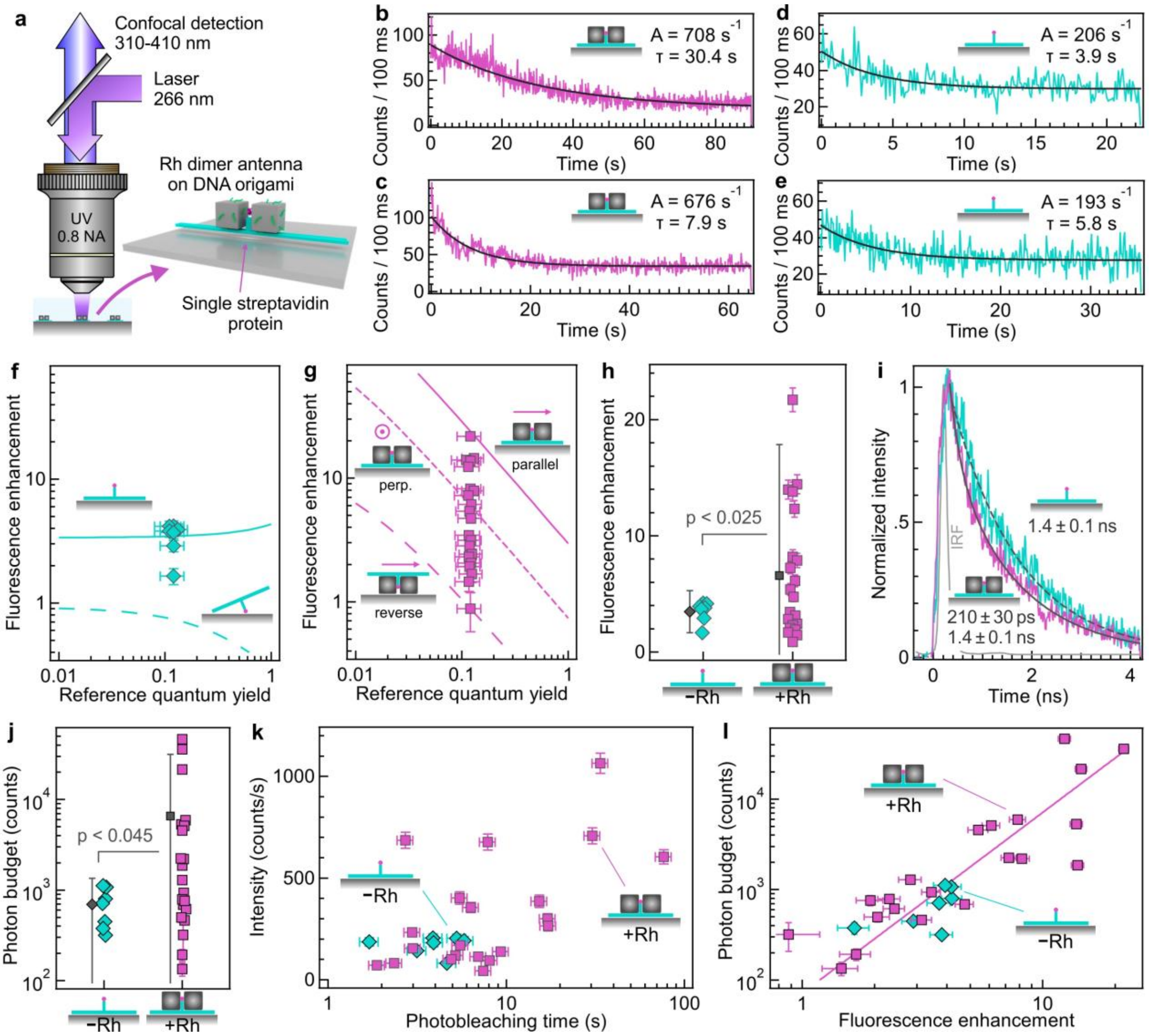

**Figure 3:** Label-free single protein detection with plasmonic-enhanced deep-UV autofluorescence. (a) Scheme of the experimental setup. (b-e) Representative autofluorescence time traces recorded for a single streptavidin protein localized on a DNA origami in presence (b, c) or absence (d, e) of the Rh NCs antenna. The autofluorescence traces are fitted with a single exponential decay (thick black line) with the respective amplitude A and decay time τ indicated on each graph. Supplementary traces are shown in the Supporting Information Fig. S6. (f, g) Fluorescence enhancement factors as a function of the reference quantum yield in homogeneous solution environment. The markers correspond to the experimental data in absence (f) or presence (g) of the Rh NCs antenna. The presentation of the experimental data include some random shift along the X-axis to better view the different points. The solid lines correspond to numerical simulations when the DNA origami base is in contact with the aluminum mirror. The dashed lines denote the simulations results where the origami position is reversed respective to the mirror, and the streptavidin is at the closest distance to the mirror. The dotted line in (g) corresponds to numerical simulations when the DNA origami base is in contact with the Al mirror and the polarization is perpendicular to the dimer main axis. (h) Comparison of the fluorescence enhancement factors in absence or presence of the Rh NCs antenna. The black markers indicate the average value with the error bars corresponding to 2× the standard

deviation. The error bars on the individual data points correspond to the experimental uncertainties, which appear on some cases smaller than the marker size. The p-values written on the graphs result from statistical T-tests performed on the distributions. (i) Normalized time-resolved decay traces measured in absence or presence of the Rh NCs antenna, cumulated over the 5 brightest systems of each case. The same data is shown in logarithmic scale in Fig. S8b. IRF indicates the instrument response function. (j) Comparison of the total photon budgets. (k) Scatter plot of the autofluorescence decay amplitude versus the photobleaching time. (l) Scatter plot of the total photon budget versus the autofluorescence enhancement factor confirming the gain brought by the brightest antennas. The line is an empirical fit following a quadratic dependence.

Figure 3 summarizes our main experimental findings monitoring the autofluorescence of single streptavidin proteins located inside the plasmonic hotspot of Rh NCs dimer antenna. The origami structures are dispersed on an aluminum mirror and scanned individually with a UV confocal microscope with 266 nm laser illumination and autofluorescence detection in the 310-410 nm range (Fig. 3a). Confocal scans in presence of the Rh nanoantennas loaded with streptavidin show distinct bright spots, which we associate to the nanoantenna signal (Supplementary Fig. S5). In absence of the streptavidin protein, no signal could be detected, confirming that the signal originated exclusively from streptavidin and indicating that the background from the Rh photoluminescence and from the DNA origami remains negligible in our case. Figure 3b-e present typical autofluorescence time traces recorded for a single streptavidin protein localized on a DNA origami in presence or absence of the Rh NCs antenna. Supplementary traces are shown in the Supporting Information Fig. S6. Contrarily to the stepwise photobleaching of a single organic fluorophore,[39] each streptavidin molecule contains 24 tryptophan and 24 tyrosine residues that bleach sequentially with overlapping decay steps, resulting in an overall exponential-like fluorescence decay. Unfortunately, the signal to noise ratio is not sufficient to unequivocally distinguish the photobleaching steps from individual amino acids. Therefore, the autofluorescence time trace of each detected streptavidin is fitted with a single exponential decay to extract the autofluorescence amplitude A and decay time $\tau$. To assess the reliability of our numerical analysis and the influence of low signal-to-noise conditions, we simulate autofluorescence traces with known amplitudes, decay times, and noise comparable to the experiments (Supplementary Fig. S7). The fitting procedure accurately retrieves the input parameters down to amplitudes of 80 counts/s, confirming the robustness of the experimental measurements reported in Fig. 3.

To promote the photostability and reduce the bleaching time, we have to lower the 266 nm laser power down to 220 nW. In such conditions, measuring the reference emission from a single streptavidin protein on a reference quartz coverslip is extremely challenging (amplitude estimated to less than 50 photons/s). Instead, we rely on the measurements recorded on the DNA origami with a single streptavidin on an Al mirror in absence of the Rh NCs dimer and compare the average amplitude with the numerical prediction of the 3.5× autofluorescence enhancement in the sole presence of the aluminum mirror (Fig. 2f). This allows us to indirectly assess the reference signal for streptavidin on a quartz substrate used to compute the fluorescence enhancement factors in Fig. 3. The simulations in Fig. 3f consider two main orientations for the DNA origami structure either lying flat on the surface (maximum 26 nm distance between the streptavidin and the Al surface) or reverse (streptavidin in direct contact with the Al metal, leading to maximum autofluorescence quenching). The influence of the quantum yield of the emitter is also taken into account as a parameter. While the quantum yield of each aromatic residue in the protein is not known, we rely on the 12% reference found for tryptophan and tyrosine,[15,77] and add a ± 3% uncertainty on the X-axis error bars together with a random ± 1% shift along the X-axis to better visualize the different data points. The good agreement between the experimental results and the numerical simulations in Fig. 3f, g validates this assumption.

In presence of the Rh NCs antenna, the UV autofluorescence is significantly more enhanced, up to a 22× as compared to the quartz reference (Fig. 3h). The experimental values show a broader dispersion, which is well represented by the variation of the DNA origami orientation respective to the Al mirror (up-front or reverse) and the incoming laser polarization (parallel or perpendicular, see Fig. 3g). Because of the relatively short photobleaching times, we could not rotate the incoming linear polarization during our experiments and had to rely on a constant random orientation for each Rh NCs antenna. The close agreement between the calculations and the experimental data in Fig. 3f, g highlights the reliability of DNA-guided assembly and its ability to precisely position both NCs and proteins. At the same time, it confirms the validity of our autofluorescence enhancement observations. Some of our earlier work used capillary-assembled Rh NCs into rectangular nanoholes to enhance the UV autofluorescence of proteins.[48] However, the described platform relied entirely on the random diffusion of the target protein in the vicinity of the nanoantenna, resulting in large fluctuations and a lack of control on the position of the emitter. Moreover, high laser power up to 30 µW were used, but the occurrence of photobleaching was not considered in this earlier work. In contrast, our approach uses DNA-guided positioning to achieve deterministic control of the location of both nanoparticles and analytes, thereby achieving a highly specific optical response with a true control on the positioning and number of target protein.

Measuring the autofluorescence lifetime from a single immobilized protein is difficult as the limited photon budget is split across the different photon delay times. By summing up the signal from 5 different proteins, we could nevertheless obtain relevant lifetime histograms in presence or absence of the Rh NCs antenna (Fig. 3i). Without the Rh antenna, the 1.4 ns autofluorescence lifetime in the vicinity of the Al mirror remains comparable to the reference value measured for a 1 µM streptavidin solution (Supplementary Fig. S8a, b), in good agreement with the numerical predictions (Supplementary Fig. S8c). Thanks to the plasmonic enhancement inside the nanogap antenna, the autofluorescence lifetime is shorter in presence of the Rh NCs. Our measurements indicate a 210 ± 30 ps component corresponding to a 6.6× lifetime reduction. The numerical simulations predict even higher lifetime reductions up to 10 or 25× depending on the DNA origami orientation (Supplementary Fig. S8c). Despite this discrepancy, the overall trend is preserved, confirming the plasmonic enhancement of the total decay rate.[39] It should also be mentioned that the spectroscopy experiments average over 24 different tryptophan and tyrosine residues with varying positions and orientations while our numerical model only accounts for a single dipole located in the center of the nanogap with a constant parallel orientation.

The photon budget defined as the total number of photons detected before photobleaching is a crucial parameter for single molecule fluorescence experiments.[78,79] For a single exponential bleaching decay, the photon budget corresponds to the product A*τ of the amplitude by the photobleaching decay time. Figure 3j compares the distributions of the photon budgets for the different detected proteins. In presence of the Rh NCs antenna, the average photon budget is increased by a factor of 9.4 from 700 to 6600. This results from the enhancement of the autofluorescence amplitude A (Fig. 3h) and also to some increase in the photobleaching time τ (Fig. 3k). As already observed for organic fluorescent dyes in the visible,[79,80] the photobleaching time follows a broad distribution covering multiple orders of magnitude (Fig. 3k). In presence of the nanoantenna, the higher excitation intensity will tend to reduce τ, yet this effect can be partially compensated by the reduced autofluorescence lifetime limiting the time spent in the higher excited state leading to photo-oxidation and bleaching.[81,82] Altogether, Figure 3l summarizes the performance comparison between DNA origami structures with and without the Rh NCs, considering both fluorescence enhancement and total photon budget. Likely due to variations in orientation relative to the laser polarization and/or the aluminum mirror, some Rh NC antennas exhibit responses similar to the reference DNA origami alone. However, a distinct population showing approximately 10× higher fluorescence intensity and 10× higher photon budget clearly demonstrates the superior performance of the UV plasmonic antennas.

**Conclusions**

In this work, we establish a complete route for transforming Rh NCs into programmable UV-plasmonic nanoantennas through surface-specific functionalization. To this end, we report an effective method for removing the native PVP ligand without compromising colloidal integrity of Rh NCs, followed by their functionalization with thiolated DNA strands using a ‘freeze-thaw’ method. Exploiting the DNA origami technique as an addressable scaffold, we demonstrate deterministic assembly of Rh NC dimers with an average interparticle distance of 10 nm and a yield of 69%. A single streptavidin protein is precisely positioned within the plasmonic hotspot, pioneering a route for single label-free single protein sensors enhanced with UV plasmonics. Importantly, owing to its addressability, the DNA origami technique introduces molecular selectivity, unlike previously reported UV-plasmonic systems in which analytes diffused randomly during the measurements.

UV autofluorescence measurements of single immobilized streptavidins show brightness enhancement factors up to 22× as compared to the reference on quartz, together with a 6.6× reduction of the autofluorescence lifetime and a 10× increase of the average total photon budget. The good agreement between our experimental results and numerical simulations confirm the validity of our conclusions and highlight the remarkable nanofabrication and positioning accuracy enabled by DNA origami templated Rh nanoantennas. Overall, this work demonstrates a versatile, reproducible, and highly specific platform for constructing UV-resonant plasmonic nanoantennas, opening new opportunities for optical studies of individual label-free biomolecules in aqueous solutions. Future optimization of nanoparticle geometry and dimer orientation promises increased performance for UV-active nanoantennas and their further applications, for example in assessing the number of tryptophan and tyrosine residues.

## Materials and methods

**Rhodium cubes synthesis:** The synthesis was carried out by the polyol method developed by Zhang et al.[63] Potassium bromide (0.45 mmol, Acros) and ethylene glycol (EG, 2 mL, J.T. Baker) were added to a 20 mL vial pre-cleaned with detergent and deionized water and dried in a vacuum oven at 80 °C. The vial, equipped with a Teflon-coated magnetic stir bar, was placed in an oil bath and heated to 160 °C for 40 min with the cap loosely fitted. In parallel, rhodium(III) chloride hydrate (0.06 mmol, $RhCl_3 \cdot xH_2O$, 38% Rh, Pressure) and polyvinylpyrrolidone (PVP, 0.225 mmol in repeating units, M.W. ≈ 55 000, Aldrich) were each dissolved in 2 mL EG at room temperature. The two solutions were then simultaneously introduced into the reaction vial using a dual-channel syringe pump at 1 mL $h^{-1}$. Following injection, the reaction was maintained at 160 °C for an additional 10 min before being cooled to room temperature.

0.4 mL of above solution was used as the seed solution and diluted with 1.6 mL EG in a cleaned 20 mL glass vial. The mixture was heated at 160 °C under stirring for 40 min with the cap loosely fitted. Separately, 0.045 mmol $RhCl_3 \cdot xH_2O$ was dissolved in 2 ml EG. 0.225 mmol PVP and 0.45 mmol KBr were co-dissolved in another 2 ml EG. These two solutions were simultaneously injected into the seed solution by a syringe pump at a rate of 1 mL $h^{-1}$. The reaction was kept at 160 °C for another 10 min and subsequently cooled to room temperature. Rh NCs of two populations of sizes with an average edge length of 24 and 40 nm were synthesized in this way, as shown in Supplementary Fig 1 a and b.

Following synthesis, the reaction mixtures were washed three times with 0.5 mL deionized water and 25 mL acetone. The obtained Rh NCs were finally dispersed in 3 mL deionized water for further treatment.

**Removal of the PVP layer on rhodium cubes:** The removal of the PVP was carried out using a modified procedure described by Zhang et al.[83] Firstly, a solution of 625 pM rhodium cubes, 0.1% SDS was centrifuged at 20,000 g for 12min to remove the PVP excess from Rh NC solution. The pellet was redispersed in Milli-Q water to reach a concentration of 3.1 nM of rhodium cubes and was briefly sonicated. At the same time a solution of 60 mg/ml of $NaBH_4$ was prepared with a 1:1 ratio of Milli-Q water and ethanol. The solution of Rh NCs and the $NaBH_4$ were mixed with the ratio of 1:1, followed by the addition of Pluronic F-127 to reach a concentration of 0.3%. The solution was then sonicated in a cold bath for 120 min. Afterwards to stop the reaction, the solution was rinsed via centrifugation at 16000g for 5 min. The supernatant was removed, and the pellet was redispersed in 0.2% SDS and briefly sonicated.

**Rhodium cubes functionalization:** 400 µL of freshly PVP-removed Rh NCs (~6 nM) were mixed with 160 µL of 200 µM T18 thiolated DNA and 12 µL of 1 M NaCl. After mixing, the solution was directly frozen at -20°C. After two hours the solution was sonicated at room temperature for 5 min to unfreeze the solution. The solution was centrifuged at 16,000 g for 5 min and the pellet was redispersed in 0.2% SDS to reach a volume of ~50 µL. Afterwards the NCs were purified from the excess of DNA using 1% gel electrophoresis at 70 V and 1x TAE 12 mM $MgCl_2$ buffer. The band containing Rh NCs was cut out and extracted from the gel.

**DNA origami synthesis:** A 7249-nucleotide-long scaffold extracted from the M13mp18 bacteriophage was folded into the desired shape using 243 staples in 1xTAE, 12 mm $MgCl_2$, pH 8 buffer. It was mixed in a 10-fold excess of staples over scaffold, and 100-fold for the functional staples (biotin, and handles). The mixture was heated to 70 °C and cooled down at a rate of 1 °C every 20 min up to 25 °C. Half of the DNA origami solution was kept for adding the streptavidin (See the **Incorporation of streptavidin to DNA origami** section). The other half was later purified by 1% agarose gel electrophoresis run at 70 V for 2 h and stored at 4 °C.

**Incorporation of streptavidin to DNA origami:** Half of the DNA origami solution from the DNA synthesis was purified from an excess of staples using 100k Amicon filters at 10000g for 5 min. Streptavidin was added to the Amicon filtered DNA origami solution to reach an excess of 20-fold and incubated for 60 min at room temperature. The mix was then purified by 1% agarose gel electrophoresis at 70 V for 2 h and stored at 4 °C.

**Verification of streptavidin binding to DNA origami:** A 1.5 nM solution of DNA origami with and without streptavidin and additionally labeled with a single fluorophore (ATTO 647N, shown as a red star in Supplementary Fig. 2a) was incubated for 3 min on a biotin-coated glass substrate. Measurements using Total Internal Reflection Fluorescence (TIRF) microscopy are shown in Supplementary Fig. 2b and 2c. Fluorescent spots detected over ~200 video frames were analyzed and summarized in Supplementary Fig. 2d. Since only fluorescence within ~200 nm from the surface is detectable in TIRF, we conclude that the observed signal comes from the DNA origami binding through biotin-streptavidin interactions, confirming successful functionalization of the DNA with a single streptavidin.

**Fabrication of the Rh NC dimers:** For the synthesis of Rh NC dimers, T18-functionalized Rh NCs were mixed with the DNA origami solution with a ratio of 20:1 in the TAE buffer containing 12mM $MgCl_2$ and 600mM NaCl. The mix was then incubated at room temperature overnight. Obtained structures were then purified by 1% agarose gel electrophoresis run at 70 V for 2 h. The additional band was then extracted from the gel. The obtained structures were used without further purification.

**Fluorescence assay of the Rh NC functionalization:** Following the workflow described in Figure 2a, two rhodium cube solutions were prepared as follows: 20 µL of OD15, 19 nm rhodium cubes were synthesized using the PVP removal protocol. One solution followed the protocol exactly, while in the other, $NaBH_4$ was omitted, keeping all other steps identical. Both solutions were then split in two and mixed with either 10 µL of 5'–AAA AAA AAA AAA AAA AAA–3' or 5'–CTC TAC CAC CTA CAT–3', each modified with ATTO 647N at the 5' end. The four mixtures were diluted in 1x TAE, 12 mM $MgCl_2$ buffer with 0.1% SDS to a final volume of 100 µL. The samples were then incubated for 2 hours at room temperature. After incubation, the solutions were centrifuged at 16,000 g for 5 minutes and pellets were redispersed in 100 µL of 1x TAE, 12 mM $MgCl_2$ with 0.1% SDS. This washing step was repeated three times to remove excess of fluorophore-modified ssDNA. Thus, the only remaining fluorophores in the solution were those attached to the ssDNA hybridized to the Rh-bound strands. Because the fluorophore is quenched when attached to the rhodium cubes, it must be released to measure its fluorescence. The melting temperature of the A18–T18 duplex in 12 mM $MgCl_2$ at 10 nM is 45 °C (according to the IDT OligoAnalyzer tool); therefore, the solutions were heated to 65 °C for 15 minutes to detach the fluorophores from the rhodium cubes. Immediately afterward, all samples were centrifuged at 16,000 g for 5 minutes at 40 °C (the maximum temperature of our centrifuge) and the supernatants were collected for subsequent fluorescence measurements.

After cooling to room temperature, the supernatants collected from the above protocol were transferred into a quartz cuvette and placed in a Duetta™ spectrometer (HORIBA Scientific). Measurements were performed using the following parameters: excitation wavelength, 647 nm; emission range, 660–900 nm; excitation band pass, 5 nm; emission band pass, 5 nm; integration time, 120 s; detector accumulations, 1; and emission increment, 0.5 nm (1 pixel). To assess the concentration of the DNA on the surface the measurements between 660 and 668 nm were averaged and compared to the fluorescence intensity of known concentration of the fluorophore-labeled DNA strands. The resulting data are shown in Figure 2b.

**Transmission Electron Microscope:** For TEM imaging, 5 µL of the solution were dropped onto EM-Tec Formvar Carbon support film on copper 300 square mesh. After 2 min the solution was removed with a paper filter, and the sample was stained using 2% Uranyl Acetate for 40 sec followed by a quick water rinsing. The measurements were then carried out using a Tecnai Spirit with an accelerating voltage of 120kV.

**UV confocal setup:** The UV microscopy experiments on single DNA origami structures were performed on a custom-built confocal microscope equipped with a LOMO 58x 0.8NA water immersion objective. A 266 nm linearly polarized picosecond laser (Picoquant LDH-P-FA-266, 70ps pulse duration, 80MHz repetition rate) was used for excitation. The laser beam was spatially filtered with a 50 µm pinhole and spectrally filtered by a short-pass filter (Semrock FF01-311/SP-25). The excitation laser was then reflected towards the sample by a dichroic mirror (Semrock FF310-Di01-25-D). The average 266 nm laser power was 0.22 µW measured before the dichroic mirror at the entrance of the microscope body. A 3-axis piezoelectric stage (Physik Instrumente P-517.3CD) was used to position the sample at the laser focus. The autofluorescence emission in the spectral range of 310 to 410 nm was collected back by the same microscope objective and separated from the excitation by the dichroic mirror and two emission filters (Semrock FF01-300/LP-25 and Semrock FF01-375/110-25). A 200 mm focal length air-spaced achromatic doublet was used as a microscope tube lens to focus the fluorescence light on a 50 µm confocal pinhole. The autofluorescence light was detected by a single photon photomultiplier tube (Picoquant PMA 175) connected to a time-resolved counting module (Picoquant Picoharp 300 with time-tagged time-resolved mode). The autofluorescence time traces were analysed with the softwares Symphotime64 (Picoquant) and IgorPro (Wavemetrics). All the UV experiments were performed in a freshly prepared 1x TAE buffer at pH 8 with 12 mM $MgCl_2$. The buffer solution was purged with argon for 10 min just before the experiments to remove the dissolved oxygen and reduce the formation of reactive oxygen species.

**Finite Element Method Simulations:** Fluorescence enhancement simulations were performed using the finite element method (FEM) in CST Studio Suite. No adaptive mesh was used. The particles were modeled as cubes with rounded edges of r = 2nm. The permittivity of water was set to 1.77, and the dielectric function of rhodium was taken from "Handbook of Optical Constants of Solids" by Edward D. Palik. The excitation light was linearly polarized either along or perpendicular to the dimer axis. The fluorescence enhancement factor was calculated by placing a Hertzian dipole at the center of the gap, oriented along the dimer axis. Then enhancement of the quantum yield of the emitter is calculated by:

$$\frac{q}{q_0} = \frac{\dfrac{\gamma_r}{\gamma_r + \gamma_{loss} + \gamma_{nr}}}{\dfrac{\gamma_{r0}}{\gamma_{r0} + \gamma_{nr0}}} = \frac{\dfrac{\frac{\gamma_r}{\gamma_{r0}}}{\frac{\gamma_r + \gamma_{loss}}{\gamma_{r0}} + \frac{\gamma_{nr}}{\gamma_{r0}}}}{q_0} = \frac{\dfrac{\frac{\gamma_r}{\gamma_{r0}}}{\frac{\gamma_r + \gamma_{loss}}{\gamma_{r0}} + \frac{\gamma_{nr0}}{\gamma_{r0}}}}{q_0} = \frac{\dfrac{\frac{\gamma_r}{\gamma_{r0}}}{\frac{\gamma_r + \gamma_{loss}}{\gamma_{r0}} + \frac{1-q_0}{q_0}}}{q_0} = \frac{\dfrac{\frac{P_r}{P_{r0}}}{\frac{P_r + P_{loss}}{P_{r0}} + \frac{1-q_0}{q_0}}}{q_0}$$

$q$ = Quantum yield of the emitter in the presence of the NCs

$q_0$= Intrinsic quantum yield of the emitter

$\gamma_r$= Radiative decay rate of the emitter with NCs

$\gamma_{r0}$ = Radiative decay rate of the emitter without NCs

$\gamma_{loss}$ = Nonradiative decay rate due to energy loss in the NCs

$\gamma_{nr}$ = Intrinsic nonradiative decay rate of the emitter with NCs

$\gamma_{nr0}$ = Intrinsic nonradiative decay rate of the emitter without NCs

$P_r$ = Radiated power of the Hertzian dipole with NCs

$P_{r0}$ = Radiated power of the Hertzian dipole without NCs

$P_{loss}$ = Power lost due to absorption in NCs

Thus, the final enhancement factor is equal to $\frac{q}{q_0}\frac{|E|^2}{|E_0|^2}\frac{\eta}{\eta_0}$, where $|E|$ is the magnitude of the electric field at the Hertzian dipole position under irradiation and $\eta$ is the collection efficiency of our optical setup.

**Acknowledgements**

G. P. A. and D. G. acknowledge funding from the European Union Program HORIZON-Pathfinder-Open: 3D-BRICKS, grant Agreement 101099125 and from the Swiss State Secretariat for Education, Research and Innovation (SERI) contract number 23.00075. M.K.V., J.W. and D.G. acknowledge funding from the European Research Executive Agency (REA) under the Marie Skłodowska-Curie Actions doctoral network program (grant agreement No 101072818). Y.Y. and Y.Z. acknowledge funding from the U.S. National Science Foundation (grant number CHE-1954838).

Supplementary Information for

# DNA-Origami-Assembled Rhodium Nanoantennas for Deep-UV Label-Free Single-Protein Detection

*Nicco Corduri[1], Malavika Kayyil Veedu[2], Yifan Yu[3], Yanqiu Zou[4], Jie Liu[3], Denis Garoli[4,5,6*], Guillermo P. Acuna[1,7*], Jérôme Wenger[2*], Karol Kołątaj[1*]*

[1] Department of Physics, University of Fribourg, Chemin du Musée 3, Fribourg CH-1700, Switzerland.

[2] Aix Marseille Univ, CNRS, Centrale Med, Institut Fresnel, AMUTech, 13013 Marseille, France.

[3] Department of Chemistry, Duke University, Durham, NC 27708, USA.

[4] College of Optical and Electronic Technology, China Jiliang University, Hangzhou 310018, China

[5] Optoelectronics, Istituto Italiano di Tecnologia, 16163 Genova, Italy.

[6] Dipartimento di Scienze e Metodi dell'Ingegneria, Università di Modena e Reggio Emilia, Via Amendola 2 Padiglione Tamburini, 42122 Reggio Emilia, Italy.

[7] Swiss National Center for Competence in Research (NCCR) Bio-inspired Materials, University of Fribourg, Chemin des Verdiers 4, CH-1700 Fribourg, Switzerland

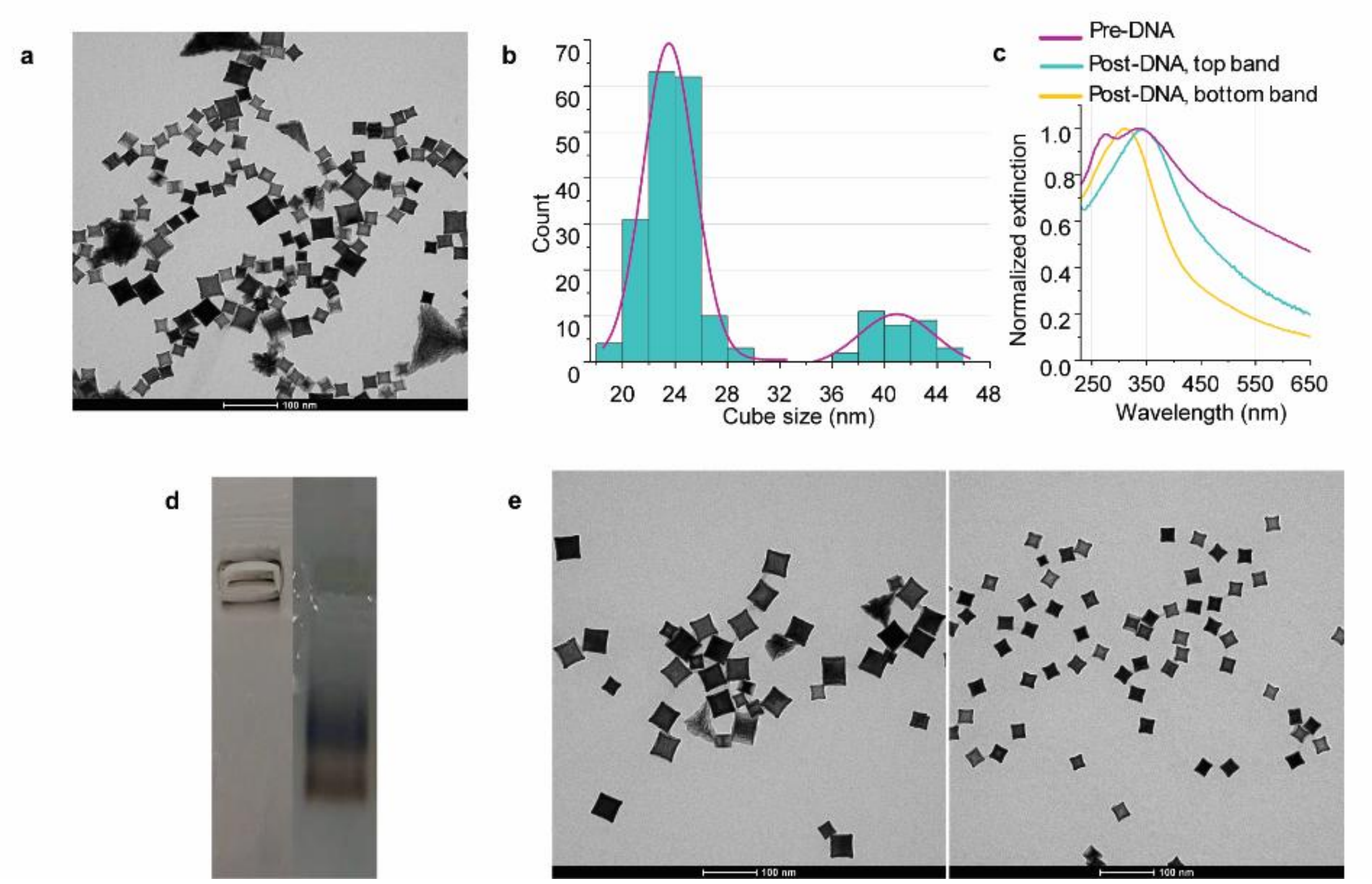

**Supplementary Figure S1:** (a) TEM image of PVP-coated rhodium nanocubes and their corresponding size distribution (b). (c) Normalized extinction spectra of 24 nm rhodium cubes. (d) Gel electrophoresis of rhodium nanoparticles functionalized with SH-T18 before and after PVP detachment (left and right respectively). The figure contains results from two different gels obtained however using the same batch of Rh nanoparticles and the same protocol. (e) TEM images of (Left) the top black band and (Right) the bottom brown band from the gel in (d).

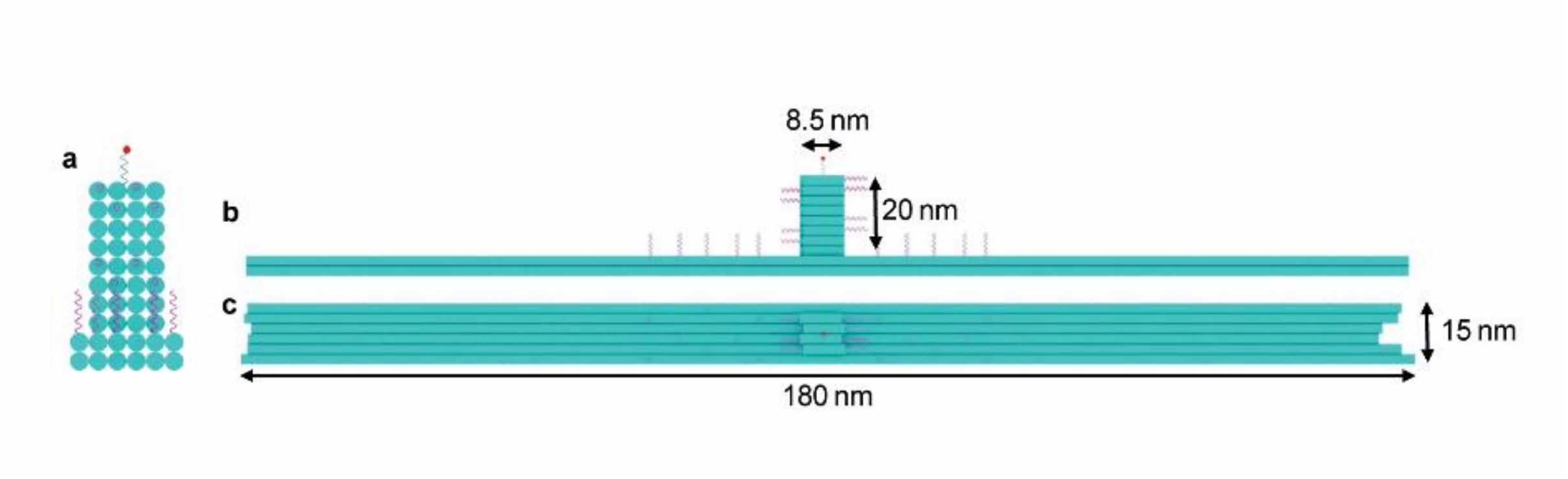

**Supplementary Figure S2:** The DNA origami design view from the (a) front, (b) side, (c) top. In pink, Poly(A)8 handles used for the binding of Rh NCs are represented by pind strands or dots, a position biotin used for the binding of a single streptavidin is represented by the red dot at the mast.

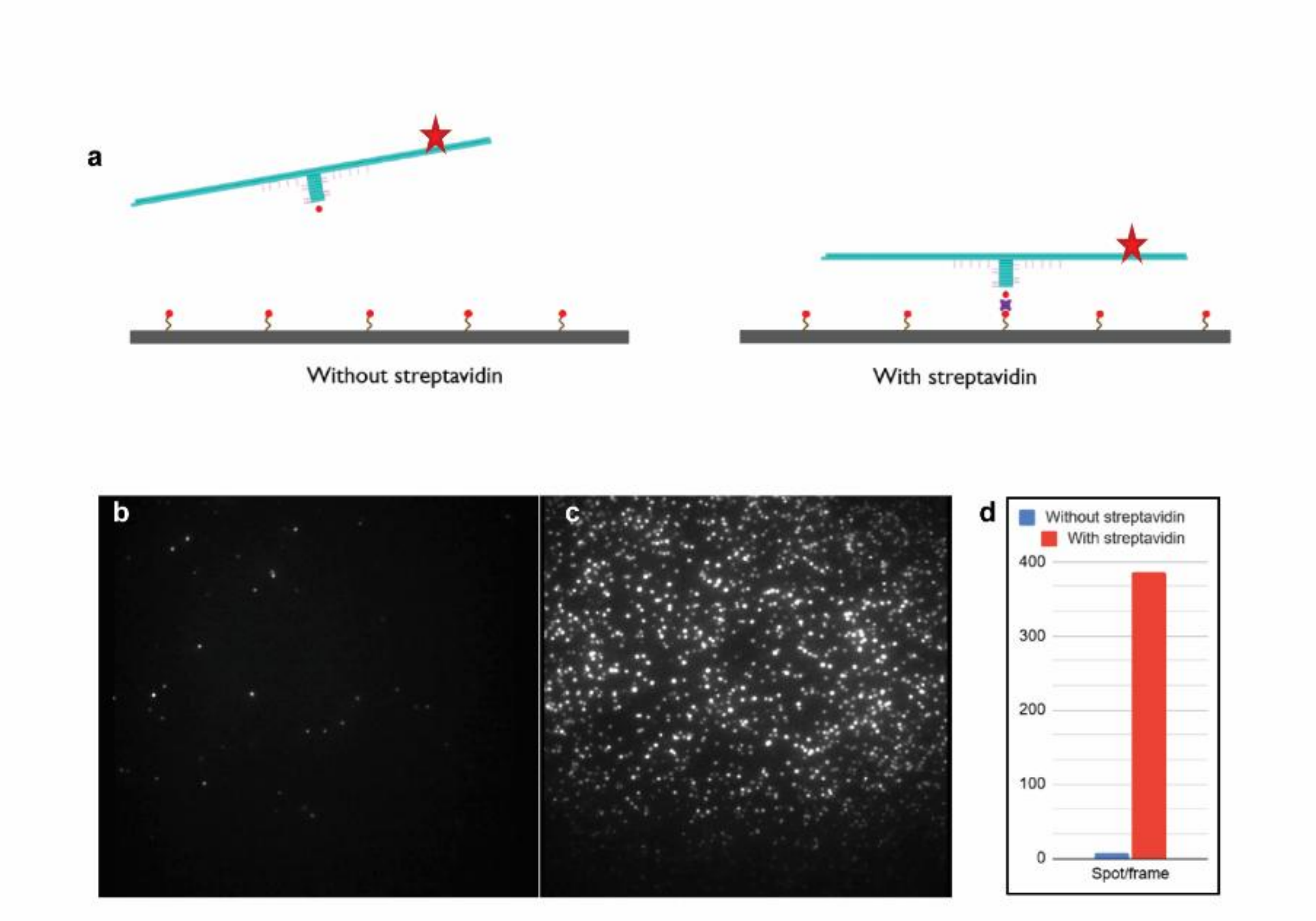

**Supplementary Figure S3:** (a) Scheme of the binding of the DNA labeled with the dye (ATTO 647N) and a single streptavidin to the biotin-coated glass surface. Fluorescence measurements of the binding of the DNA origami (b) with and (c) without streptavidin. (d) Number of detected spots over ~200 frames in the absence (blue) and presence (red) of streptavidin

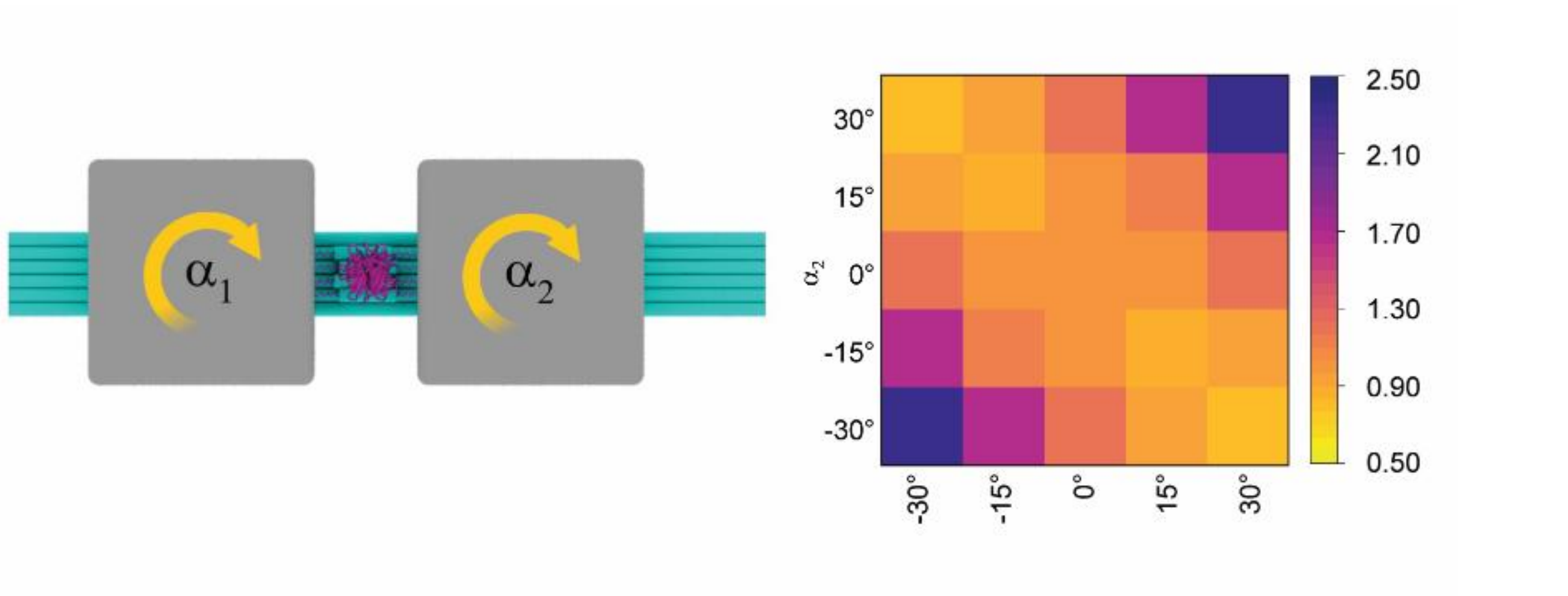

**Supplementary Figure S4:** Heatmap of the FE at the streptavidin position as a function of the orientation of the rhodium cubes on the DNA origami. The results were normalized to the FE at $\alpha_1$ = 0° and $\alpha_2$ = 0°.

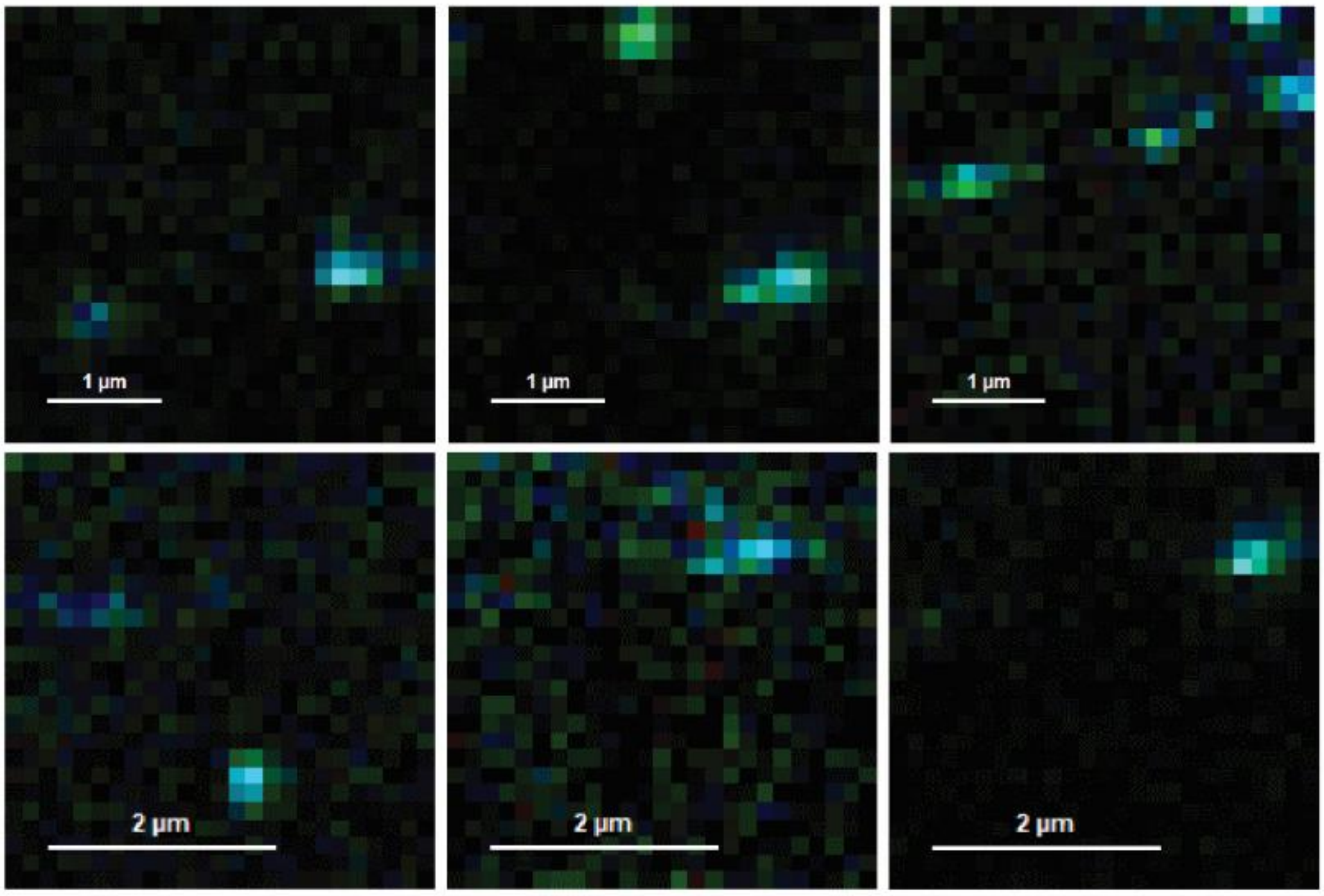

**Supplementary Figure S5:** Confocal scan of an aluminum surface coated with 24 nm rhodium cube dimers containing streptavidin in a 10 nm gap. No signal was obtained in absence of the streptaviding protein.

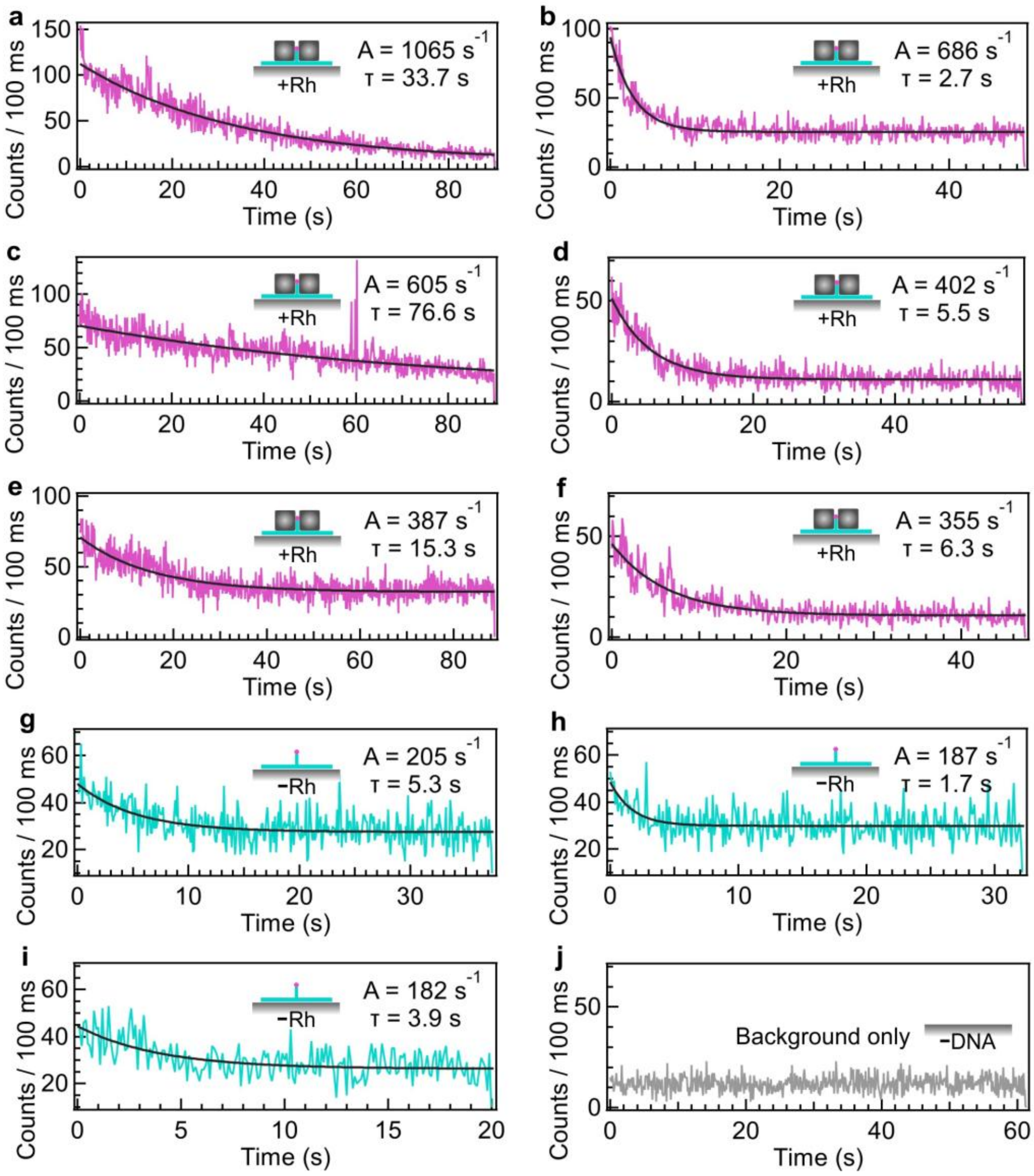

**Supplementary Figure S6:** Representative autofluorescence time traces recorded for a single streptavidin protein localized on a DNA origami in presence (a-f) or absence (g-i) of the Rh NCs antenna. The autofluorescence traces are fitted with a single exponential decay (thick black line) with the respective amplitude and decay time indicated on each graph. The case (j) correspond to the background measured in absence of the antenna.

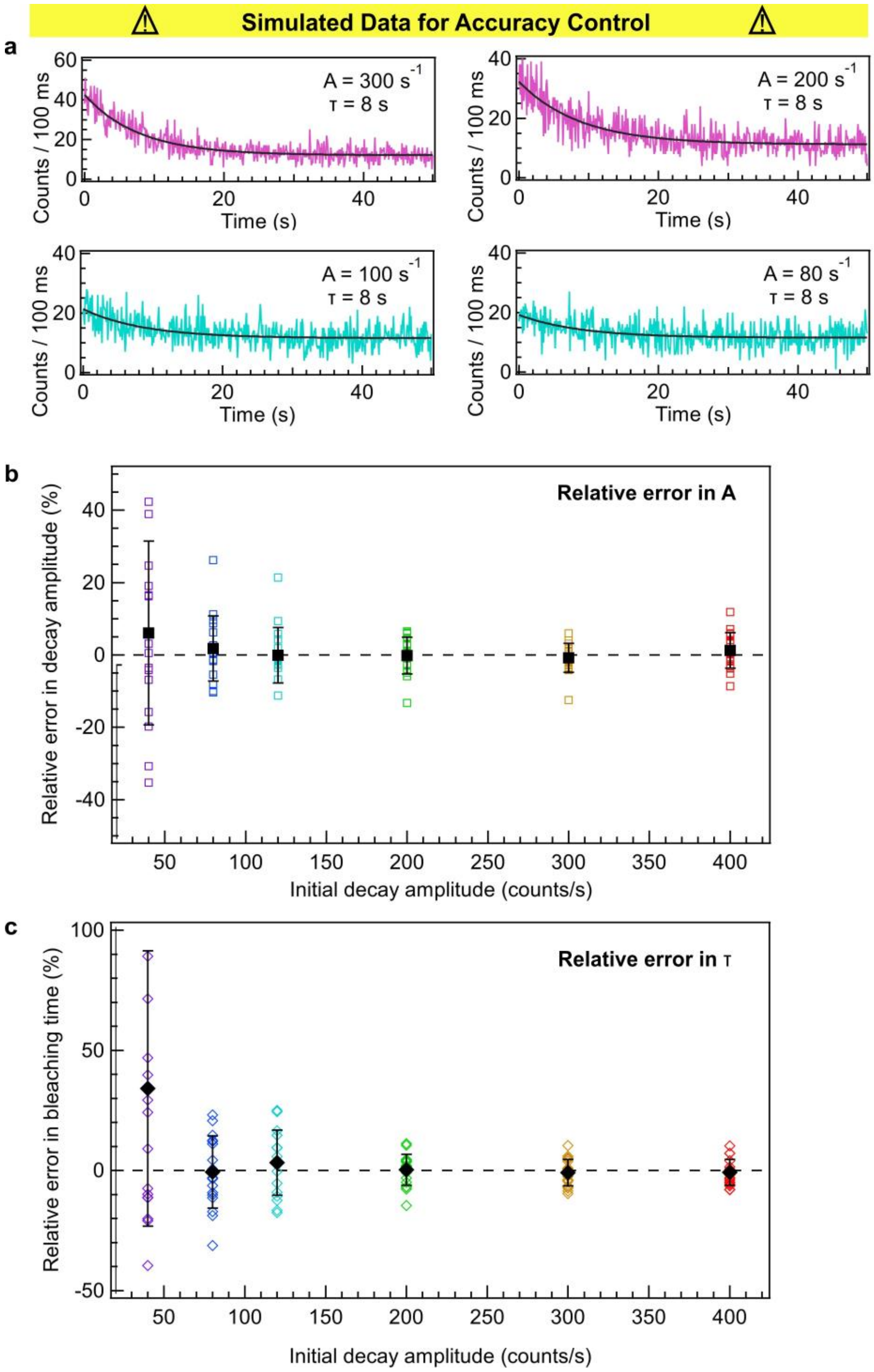

**Supplementary Figure S7:** Control of our data analysis in conditions of low signal to noise. (a) Here the autofluorescence traces have been simulated with amplitudes, decay times and noise similar to the experimental data (Fig. 3 and S5). Our aim is to check the fit results when the initial ground-truth parameters (amplitude & decay time) are perfectly known. The relative errors on the amplitude A (b) and decay time τ (c) can then be plotted as a function of the initial amplitude input. The initial value of the decay time is found to have a marginal influence, we use a value of 8 s corresponding to the mean of the experimental values. In (b,c) the black markers indicate the average value out of 20 individual trials, the error bars correspond to one standard deviation. Down to an amplitude of 80 counts/s, the average of the relative error converge towards zero, indicating a good accuracy for our measurements. In our experiments, only two datasets gave

results below this value (43 and 71 counts/s respectively), all the other cases where above 80 counts/s.

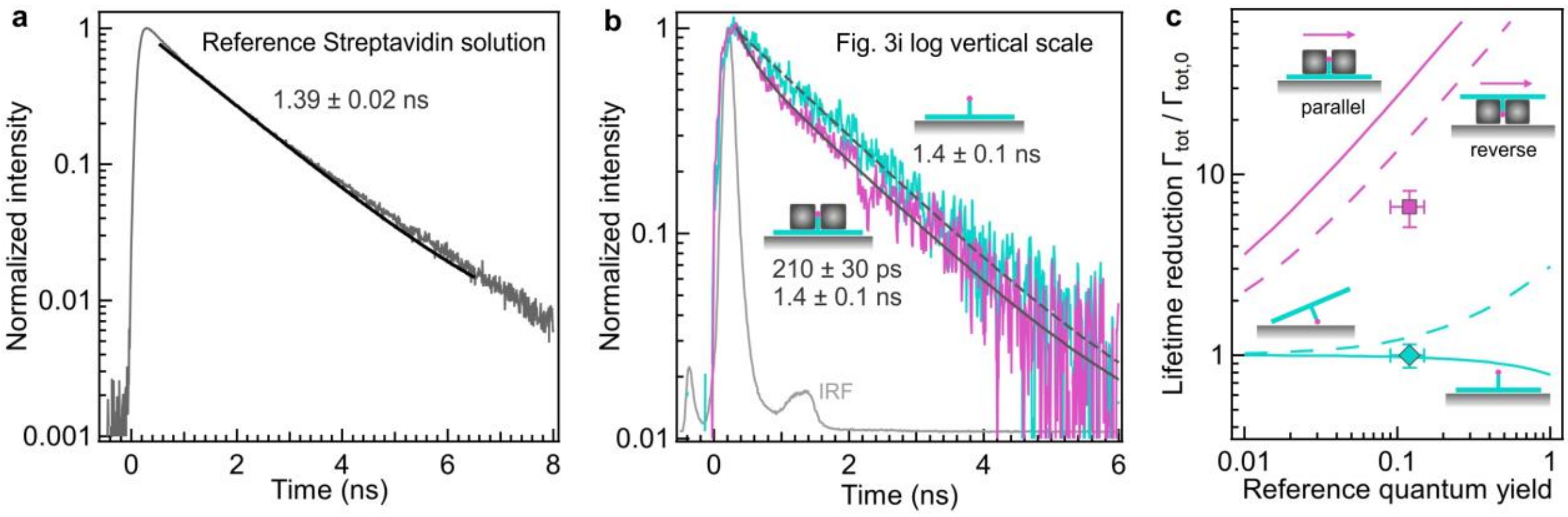

**Supplementary Figure S8:** (a) Normalized time-resolved decay traces for a reference streptavidin solution at 1 µM concentration with 0.5 µW excitation power at 266 nm. The major lifetime component at 1.39 ns corresponds to the lifetime measurement in absence of the Rh NCs. (b) Same as Fig. 3i with a logarithmic vertical scale to see better the long lifetime component. (c) Fluorescence lifetime reduction (equivalent to the total decay rate constant enhancement) as a function of the reference quantum yield in homogeneous solution environment. The markers correspond to the experimental data in absence (diamond marker) or presence (square) of the Rh NCs antenna. As in Fig. 3f,g, the solid lines correspond to numerical simulations when the DNA origami base is in contact with the aluminum mirror. The dashed lines denote the simulations results where the origami position is reversed respective to the mirror, and the streptavidin is in closest distance to the mirror.

| Name | Sequence |
|---|---|
| Core1 | AATATCGCATCATACAAGATTTAGATCATATG |
| Core2 | GGGCGAATTTGCCCCTTCACCGCCGTTGCG |
| Core3 | AGGCCGCTGAACGGTCGGAGATTTGTATCAT |
| Core4 | CAACAGCTCCAAATCAGGACTCCAACGTCAAA |
| Core5 | ACCAGACCTACTAATAAATGGTCAGATGAACG |
| Core6 | GCGGGGAGAGTGTTTTGGCCTTGCCTGTGTG |
| Core7 | AGCCGGCGTAGACAGGCTATGGTTCTGAGTAG |
| Core8 | TCCAATAAGTTTTAATCTGAATATCAGTTGAG |
| Core9 | ATGCGATTATACCACATTTACCAGCAAAGCGA |
| Core10 | GCCGCCACTCATCAATAGCACCGTAATCAGTA |
| Core11 | TCTAGAGGGGAAGCATCAGTCGGGGAGACGGG |
| Core12 | TTAGAGCCGTCAATACGCGTCTGGCCTTCCT |
| Core13 | AGCGAAAGAACCGAACGTCGAAATCCGCGACC |
| Core14 | CACTATTAGGTGGTTCTCACCAGTAAACCTGT |
| Core15 | GGTCAGTATCTTTAATAGCGTAAGTTATTTAC |
| Core16 | AAACAGGGAAGAAAAGGCCGAACAGTTTATTTTGTC |
| Core17 | TTCGCCATAAAGGAATGATTCTCCGTGGGAA |
| Core18 | TTATTTATAAAAGAACATAATAAGGGAAGGTA |
| Core19 | TTTGCGGACTTGATATCAAAAGGGCCTATTAT |
| Core20 | CCTGAGAAGGCGGTTAATAGCCCAATGCGCC |
| Core21 | GAGTAACAAGTTTGAGACCAGGCATGCAAGGC |
| Core22 | ATCAATTCGGAAGCAATCATTTTATGCAGAT |
| Core23 | CTTGAAAACGCAGTATTATTCATTAAAACAAA |
| Core25 | AAAAACCAAAGCTGCTTGGCTCATTATACCAG |
| Core26 | ACCATCACGATTGCCCAGCAGGCGAAGCCTGG |
| Core27 | CCAGCATTCATCATATTGCCTTTAGCGTCAGA |
| Core28 | GGCGATCGGACATTCTTAAAAATAGATAGGTC |
| Core29 | GTCACCAGCGCCACCCTTAGGATTAGCCGCCG |
| Core30 | ATAGGCTGTCGCTGAGAAGAGGCAAAGCCCAA |
| Core31 | CAGCTGGCCTGCATTAACAATTCCGCAAAATC |
| Core32 | TTTAACGTCAGATGAAACATCAAGAAAGGTGA |
| Core33 | CAGTTTCACATGAAAGAGAGCCACAGCGTTTG |
| Core34 | ACAGGAAGATTGTCCCCCTTATTCACCCTCATTTGTTTC |
| Core35 | TTTTGTTTAACCGAGGAGCTATCTCAAAAGGG |
| Core36 | ACGGTGTCTGGAATCAGGGATAGCAAAGAATA |
| Core37 | ATTAATTTCAGTCAAACAGAAAAGGATTCCC |
| Core39 | CTAATTTGATAAATAATGAGAATCTAGCAAGC |
| Core41 | TCGAGAACGAGGCGTTCAATTTTAAAACGATT |
| Core42 | CAGAACGCGCCTGTTTGCGCCCAAGCCATATT |
| Core43 | TTCTGACCGAGTCAATCTACCTTTTTGAATTA |
| Core44 | AGCATCACGACCTGAAGCGCGAACCATTGCAA |
| Core45 | AATACATAAAAGAAATACTTGAGCCATTTGGG |
| Core46 | AGTTGCCGAGGAATTAAGCAAGCGGTC |
| Core47 | GTAGCCAGGGAAGATCCAGCTTTTTTCCCAG |
| Core48 | ATTTTCGACTTACCAGATTACTAGATTTCATC |
| Core49 | CTTACCAAAAGAAAACGTTTTTATTTTCAT |
| Core50 | ACCTACCAAAGACACCGCCAAAGATACCGAAG |
| Core51 | ATTTTGCAGCATGTAGCCAAGTACCGCACTCA |
| Core52 | AATTAGAGAACCGATTGGCAACTAATAACG |
| Core53 | GTTGATAGATATAAGCATAAGTATAGC |
| Core54 | AGAGTACTCACGCTAACCTTTAATTGC |
| Core55 | ACCGAGTAATTCACCACCGCCAGCTGATAGCC |
| Core56 | GGCTGTCTTTCAACGAGCCGGCTTATC |
| Core57 | ATATTCGGGCTGACCTAGCGATTATACCAAGC |
| Core58 | TCCTGAACCGCTAACGGCTTATCCGGGTAAT |
| Core60 | GCTTCTGGTTAGGAGCAACATTAAATGTGAGC |
| Core61 | TCGTCGCTTCGGGAGAAATTACCTGAGCAAAA |
| Core62 | CTTGCCCTACCCTCGTTCAACTATGCGGATG |
| Core64 | CGATTGGCACAAAGATCGGCATTTTCGGTCATAGATAAG |
| Core65 | ACGCGAGACTGAGAGAAGTGAATTTCTGTAAA |
| Core66 | TCCTTTTGTATTTTCAGAGCTGAAATCTACAAAGGCT |
| Core67 | AGCAAAATAAGAGGAAATATGCAGGAACAAC |
| Core68 | CCGGAATACAGCCCTCTAACGATCGGTAGCAACGGCT |
| Core69 | ATTGGCAGAAAGAGTCTCACTTGCGCTTTGAC |
| Core70 | CCCTTTTTAAGCGCATAAATGAAATAGTTGCT |
| Core71 | AAAAGATTTAAGCAATTTTCGGAATGAGAAAG |
| Core72 | CACTAAAACACTCACGAACTAACACTAAAGT |
| Core73 | TGGTTTAATTTCACCAAAAGGCATAGTAAAGGTCAGGATTAG |
| Core74 | GGAATTATGACAGGAGAAAATCACGCCACCCT |
| Core75 | GGATTATACTTCTGAACCGGAAACGTCACCAA |
| Core76 | ACAATCCTCAGCCACCAATAGAAAATT |
| Core77 | ACAGAAATCATAAAGGTGAGGGAGAGCAAGAA |
| Core78 | TCTGAAAGCGGAGTGAAAGCGGATAACCGAT |
| Core79 | ACAATGAACCTGAACAAAATAAGTCCTGAAT |
| Core80 | TGCTCCATTTTTCATGGGAACGAGTAAAGTTT |
| Core81 | CCCTCAATCAATATCTCGTAATGGCCGAACGA |
| Core82 | TTACCATTTTACCAGCACGGAATAAAAGTTAC |
| Core83 | GTAAATTGGGCGGGCGTATCATTGAGA |
| Core84 | GAAACGCATATCAAAATCACCAGTAGCACCA |
| Core85 | CGTGCCAGGAAAGGGGGTACCGAGTAACCGTG |
| Core86 | AGGGATTTAACGTGGCAGCGGTCACGCTGCGC |
| Core87 | GGTGCCTAAAGCTTGCGCCAGGGTCCGGCACC |
| Core88 | CTGAACACATAGCAATAAACGCAAATATAAAA |
| Core89 | CTTTTACAATTAATTAAATTTCATTTAACCTC |
| Core90 | CAGAAGGAAACGTCAATAGACGGGCCTCCCGA |
| Core91 | ACCGTGTGCCAGTTACAGAGATAAATATGTAA |
| Core92 | AGGGTGGTAAAGCACTGTTTGGAACAAGAGTC |
| Core93 | ATCAGGCCAGCGCACTTCATTGCCTGA |
| Core94 | CAGGAAAATAATAACATGTCCATCCCGATTAA |
| Core95 | ACCCTTCTCTTGCTGAAAACAGAGGTGAGGC |
| Core96 | TTTGCTCATTCCAGACACAACGCCTCCATTAA |
| core97 | GGTGCCGTTTTTCTTTCGAAATCGACACAACA |
| core98 | TGAAACCACAGAACCGAGCCACCATGCCGTCGAGAGG |
| core99 | GCCGGAGATAAAAGTTTAAATTGTAAACGTTA |
| core100 | TGACAAGAGCCAGAGGCCCACGCATTGCATCA |
| core101 | GGGAGCCCCCGATTTACCGCGCTTGAGATAGG |
| core102 | TATTCAACTATTAAATCGCATTAAATTTTTG |
| core104 | CATAAGGGACAGCATCAGGAAGTTTGTAGCAT |
| core105 | TCACGACGTTGGGCGCTTTGGTAAAAC |
| core106 | ATTCCATATAACAGTTCCCCAAAA |
| core107 | GAATACCCCCCAATCCAAGTCAGAGGTATTCT |
| core108 | CAGAGATAGCGATAGTGAATAACATAA |
| core109 | TCTGTATGGAGAAGGATCAGAACCCGGAACCA |
| core110 | CCTTATTACATAGCGAATATAACTCCCACAAG |
| core111 | ACGTAACAAAATAGCGAGATTCATAATGCTGT |
| core112 | ATAAAAGGGTGCGGGCAGCTGTTTCTGGTAAT |
| core113 | AAACACCGAATAAACATAATTTAGGCAGAGGC |
| core114 | GACGGCCAGTGCCATGAGTGAATTAATTGCTGGCCCTGAGAG |
| core115 | TGTCGTCTGTACCAGGCACCGTACTCAGAGCC |
| core117 | ACAGAGTTAAAAAGCCGCTTTGAGGAC |
| core118 | TACGAGCCATCCCCGGGATGTGCAAGCGCCA |
| core119 | AGGACAGATTTTGCGGAATGCCACTGAGTTTC |
| core120 | GCTTAGAGATTTCGCAGTAGTAGTTAATGCC |
| core121 | GATTAAGTCCGCTTTCAAAGTGTAAAAATCCT |
| core122 | ATTTACGACCCAGCTATTAGCGAAAGAATTAA |
| core123 | GCTGAGAATAAATTTAAATCGCAACAACGCTC |
| core124 | GAAGTTTTACCGGATAAAAAATCTACGTTAATAAAATCTT |
| core125 | ATGGCAATCAGAACCCGCCTCCCTCAGGAGG |
| core126 | CTATCATAGACGAGAACATTGTGAATTACCTT |
| core127 | ATTCCAAGAACGGAGGTTTTGTCAAGATATAGCAGCCTTTA |
| core128 | CTCACTGCTGGGTAACATGCCTGCGACAGTAT |
| core129 | AACAGTTGTCAGGCTGGCGCATCGCTCGAAT |
| core130 | TTAAATCAAAAACTACTGATAAACATTAACA |
| core131 | ACAACTCGCGTTCTAGGCATGTCATTTGACCA |
| core132 | ACGACGACGAATCATATATAAAGCGACAAAGA |
| core133 | ATTTAGAATAGCTATTAGAGAATCATAACCTG |
| core134 | CTCCTCAAGGATTTTGTAACACTACGAAGG |
| core135 | GGAGAGGGGTATTAGATTTAACCAATAGGAAC |
| core136 | AAATATCTTGCCGGAAGGGACGACAGGTCGAC |
| EC1 | CCCCCAGAGCGGGAGCTAACTTTCCTCGTTAGAATCCCC |
| EC2 | CCCCAGCGGGCGCTAGGGCGCTGATAA |
| EC3 | CCCCATTTTTGAATG |
| EC4 | CCCCTACCGACAAAAGGTAATAAGAGAATATAAAGCCCC |
| EC5 | CCCCGCAATACTTCTTTGATTAGACGC |
| EC6 | CCCCTGACGCTCAATCGTCTGGAAATACCTACATTTCCCC |
| EC7 | AGCGAAAGGCCCC |
| EC8 | CGTGACAGGAGGACGCAAATTAACCGTTGTACCCC |
| EC9 | CCCCCATATGCGTTA |
| EC10 | CCTGATTGGTGAATAACAGTACATAAATCAATCCCC |
| EC11 | CCCCAGTTACAAAATCGCGCAGAGGCGAATTAAATG |
| EC12 | CCCCATATGTGACTTTGAATACCACCCC |
| EC13 | CCCCCGCTGAGAGCCAGCAGGCCTGCAACAGTGCCACCCC |
| EC14 | CCCCTATATTTTAGTTAAAAAAGCCTGTTTAGTATCCCC |
| EC15 | GAAACCTTGCTTATCAAAACCCC |
| EC16 | CCCCTCATAGGTAAACTTTTTCAAACCCC |
| EC17 | TCATGAAATGGAAATACGTGGCACAGACAATCCCC |
| Mittle1 | CCCCGACCCTGTAATACTTTGAATGCCCC |
| Mittle2 | CCCCGAGTGTACAAACAGTTAAGGACCCC |
| Mittle3 | CCCCCGGTTGCTTTTGAAAGCCTTCCCC |
| Mittle6 | CCCCTCCTCTAAAAATTCGGGGTCACCCC |
| Mittle7 | CCCCTATTATAGCCTGCCTAAAAGCCCCC |
| Mittle8 | CCCCAATGCTTTTGGTAATATAAATCCCC |
| Mittle14 | CCCCAGAAAAGGTCTTTCGCCGACACCCC |
| Mittle16 | CCCCAGCCATGCGGGAGTGATACAGCCCC |
| Mittle17 | CCCCAACTACAGAAAACGAATTTCTCCCC |
| Mittle21 | CCCCGTAAGCGTAAATATTGAATACCCC |
| Mittle22 | CCCCTAAACAGCTTGATACCAAAATCCCC |
| Mittle24 | CCCCTGTGTAGGTAAAGATTTCACACCCC |
| LHF1 | TACAAATTGCCAGTAAAGTAATTCTGTCCAG |
| LHF3 | GCCACCCTTCGATAGCATAATCCTGATTGTTT |
| LHF4 | CACCAACCAAGTACAAGTACAGACCAGGCGC |
| LHF5 | CGACATTCCCAGCAAAATTATTTGCACGTAAA |
| LHF6 | CCTTTTTTTTCATTTCAACAATAACGGATTCG |
| LHF7 | ACGGGTAATAAATTGTTGACCAACTTTGAAAG |
| LHF8 | ATTAATTATGAAACAATATACAGTAACAGTAC |
| LHF9 | AATATTGACGTCACCGTGCGTAGATTTTCAGG |
| LHF10 | AACAGTAGCCAACATGACATGTTCAGCTAATG |
| LHF11 | AAGAACGCAAGCAAGCATAATATCCCATCCTA |
| LHF12 | CTTGCGGGGTATTAAAAAACCAATCAATAATC |
| LHF13 | AAATCAGATCATTACCATCAACAATAGATAAG |
| LHF14 | CATATGGTAGCAAGGTAATGGAAGGGTTAGA |
| LHF15 | CCATCTTTCGTTTTCAAACCACCAGAAGGAGC |
| LHF16 | GAGCCACCATCAAGTTTCCTGATTATCAGATG |
| RHF1 | GAGCACGTGCAAGTGTGAGAAAGGAAGGGAAGAA |
| RHF2 | CATCTGCCACCCGTCGTGAGGAAGGTTATCTA |
| RHF3 | ATTATTACTTGGGAAGTTCATTACCCAAATCA |
| RHF4 | GTTTGATAAGAACGTAGTTTTTTGGGGTCGA |
| RHF7 | GTAATCGTGCTCATTTCTTTACAAACAATTCG |
| RHF8 | CACGCTGGAAACCGTCATGGCCCACTACGTGA |
| RHF9 | ACGTTGGTGGATTGACGGTCAGTTGGCAAATC |
| RHF10 | ACATAACGACTTTAATACACCAGAACGAGTA |
| RHF11 | CCTTATAATTGTTCCAAAATCGGAACCCTAAA |
| RHF12 | GCTACAGCACACCCGGAGCTTGACGGGGAA |
| RHF13 | GCTATTAGTTAACACCCAAATGAAAAATCTAA |
| RHF14 | CTAAAACACAGAAGATAACCTCAAATATCAAA |
| RHF16 | CGGCCTCACTTTCATCACTAACAACTAATAGA |
| CSF3 | CCCCATAATTTTTTCACGTTGAGGACCCC |
| CSF4 | CCCCAATACTGCAAATCTCCAAAAACCCC |
| PLF1 | TGAGCGCTTAAGCCCATGGCATGATTAAGACT |
| PLF2 | CGGCTTAGGCAAATCCATGGTTTGAAATACCG |
| PLF3 | ATGCTGATGTTGGGTTTAGCTTAGATTAAGAC |
| PLF4 | TAACAACGGGCTTAATGGCGTTAAATAAGAAT |
| PLF5 | GAAGATGACATTTAACATTTTCCCTTAGAATC |
| PLF6 | AATTGAGTAATATCAGAAAATAAACAGCCATA |
| PLF7 | ATTATCACCGGAAATGTTAGCAAACGTAGAA |
| PLF8 | CGTAGGAATATAGAAGAGCGTCTTTCCAGAGC |
| PCF3 | TAGGAACCCCACCCTCTATTAAGAGGCTGAGA |
| PCF5 | TGACCCCCTCATCAAGAGTAATCT |
| PCF6 | AATTCTGCATGTTTTAAGCCCGAAAGACTTCA |
| PCF7 | CAAATATTTGAGTAACATTATCAT |
| PCF9 | TCAGGACGAGGTAGAAAGAGGCTTTTGCAAAA |
| PRF1 | ATCCAGAAAAACTATCTATAATCAGTGAGGCC |
| PRF2 | TCGTAATCATCCGCTCATGAATCGGCCAACGC |
| PRF3 | ACCACCAGTCGCCATGGCCAACAGAGATAGA |
| PRF4 | AAGAACTCCAATATTAGTCACACGACCAGTA |
| PRF5 | CAAACGGCGTAGATGGCGCAACTGTTGGGAAG |
| PRF6 | GTTGAGTGATCAAAAGTGCGTATTGGGCGCC |
| PRF7 | AAATTGTTATGGTCATCTCTTCGCTATTACGC |
| PRF8 | GTAACCACGGCGCGTAAACGGTACGCCAGAAT |
| Floor Left 1 | GAGCCACCATCAAGTTTCCTGATTATCAGATGAAAAAAAAAA |
| Floor Left 2 | CTGTAGCGTCATAATCGTTGAGGCAGGTCAGAAAAAAAAAAA |
| Floor Left 3 | TCCACAGAGGTGTATCGGATAAGCCCTCAGAAAAAAAAAAA |
| Floor Left 4 | CAGAACCGCATGTACCGCTAAACAACTTTCAAAAAAAAAAAA |
| Floor Left 5 | CACCAACCAAGTACAAGTACAGACCAGGCGCAAAAAAAAAA |
| Floor Left 6 | TAAAGACTGTTACTTAGGCGCAGACGGTCAATAAAAAAAAAA |
| Floor Left 7 | TTTAGTACTACAAACTGTTAGTAACCCTCAGCAAAAAAAAAA |
| Floor Left 8 | TGCTCCATTTTTCATGGGAACGAGTAAAGTTTAAAAAAAAAA |
| Floor Right 1 | GAGTCTGGAAATAATTGATAATACATTTGAGGAAAAAAAAAAA |
| Floor Right 2 | GCCATCAAAGCAAACATTTGAGAGAAGGTGGCAAAAAAAAAA |
| Floor Right 3 | GTAATCGTGCTCATTTCTTTACAAACAATTCGAAAAAAAAAA |
| Floor Right 4 | ATATTTTGTTGATAATTCACCATCAATATGAAAAAAAAAAA |
| Floor Right 5 | AATTCTGCATGTTTTAAGCCCGAAAGACTTCAAAAAAAAAAA |
| Floor Right 6 | ATTATTACTTGGGAAGTTCATTACCCAAATCAAAAAAAAAAA |
| Floor Right 7 | TTAGATACCTTAATTGTCGAGCTTACGACGATAAAAAAAAAA |
| Floor Right 8 | TTTAGCTAATAAGAGGACTCCAACGAGCAACAAAAAAAAAA |
| Mast Left 1 | AAAAAAAAAACAATGCAATGCTAAACAGT |
| Mast Left 2 | AAAAAAAAGTAAGCGTAAATATTGAATA |
| Mast Left 3 | AAAAAAAAGTGCCTTGCAAAAATCGGAAC |
| Mast Left 4 | AAAAAAAAAGAAAAGGTCTTTCGCCGACA |
| Mast Left 5 | AAAAAAAATAAACAGCTTGATACCAAAAT |
| Mast Left 6 | AAAAAAAAAGCCATGCGGGAGTGATACAG |
| Mast Left 7 | AAAAAAAAAAAGGCTCCAAAAGGACGTCC |
| Mast Left 8 | AAAAAAAAATTGCCATTGAATTTGTATC |
| Mast Right 1 | AAAAAAAATGTGTAGGTAAAGATTTCACA |
| Mast Right 2 | AAAAAAAAGTTTATTCGAGGTGAGAATGA |
| Mast Right 3 | AAAAAAAACGGTTGCTTTTGAAAGCCTT |
| Mast Right 4 | AAAAAAAATATTTCAACGCAAGGAATTAA |
| Mast Right 5 | AAAAAAAACTCAGTGCCCGTACTGAGTAA |
| Mast Right 6 | AAAAAAAAAATACTGCAAATCTCCAAAAA |
| Mast Right 7 | AAAAAAAATATTATAGCCTGCCTAAAAGC |
| Mast Right 8 | AAAAAAAAAATGCTTTTGGTAATATAAAT |
| Biotin R | ATCGTCATCATACATGCATTTTACCGTTCCATTTT - Biotin PEG |
| Non Biotin R | AAAAAGTACC |

**Supplementary Table S1:** List of all DNA sequences used to fold the DNA origami structure for the Rh dimer. All sequences were purchased from IDT (Integrated DNA Technologies), except for "Biotin R" and "Non biotin R," which were obtained from Biomers GmbH.